\documentclass[12pt,preprint]{aastex}

\newcommand{\msun}{$M_{\odot}$}
\newcommand{\vmon}{A0620-00}

\begin{document}
\title{Near-Infrared Spectra of the Black Hole X-ray Binary \vmon}
\author{Cynthia S. Froning\footnote{Visiting Astronomer at the Infrared Telescope Facility, which is operated by the University of Hawaii under Cooperative Agreement no. NCC 5-538 with the National Aeronautics and Space Administration, Science Mission Directorate, Planetary Astronomy Program.}}
\email{cfroning@casa.colorado.edu}
\affil{Center for Astrophysics and Space Astronomy, University of Colorado, \\
  593 UCB, Boulder, CO 80309-0593}
\author{Edward L. Robinson}
\email{elr@astro.as.utexas.edu}
\affil{Department of Astronomy, University of Texas at Austin, 
Austin, TX 78712}
\and
\author{Martin A. Bitner}
\email{mbitner@astro.as.utexas.edu}
\affil{Department of Astronomy, University of Texas at Austin, Austin, TX 78712}

\begin{abstract}

We present broadband NIR spectra of \vmon\ obtained with SpeX on the IRTF.  The spectrum is characterized by a blue continuum on which are superimposed broad emission lines of \ion{H}{1} and \ion{He}{2} and a host of narrower absorption lines of neutral metals and molecules.  Spectral type standard star spectra scaled to the dereddened spectrum of \vmon\ in K exceed the \vmon\ spectrum in J and H for all stars of spectral type K7V or earlier, demonstrating that the donor star, unless later than K7V, cannot be the sole NIR flux source in \vmon.  In addition, the atomic absorption lines in the K3V spectrum are too weak with respect to those of \vmon\ even at 100\% donor star contribution, restricting the spectral type of the donor star in \vmon\ to later than K3V. Comparison of the \vmon\ spectrum to scaled K star spectra indicates that the CO absorption features are significantly weaker in \vmon\ than in field dwarf stars. Fits of scaled model spectra of a Roche lobe-filling donor star to the spectrum of \vmon\ show that the best match to the CO absorption lines is obtained when the C abundance is reduced to [C/H] = -1.5.  The donor star contribution in the H waveband is determined to be $82\pm2$\%.  Combined with previous published results from Froning \& Robinson (2001) and Marsh et al.\ (1994), this gives a precise mass for the black hole in \vmon\ of $M_{1} = 9.7\pm0.6$~M$_{\odot}$. 

\end{abstract}


\section{Introduction} \label{sec_intro}

Among X-ray binary systems (XRBs), 18 or more have been identified as containing probable black hole (BH) accretors \citep{mcclintock2005}.  
The BH masses measured to date appear to fall into a limited range.  From a Bayesian analysis of the observational parameters of several low-mass XRBs, \citet{bailyn1998} concluded that 6 of the 7 systems measured had BH masses clustered around 7~\msun and that the overall population was heavily biased away from low BH masses (3 -- 5~\msun). Since that study, at least one system, J0422$+$32, has been found to have a measured mass of about 4~\msun \citep{gelino2003}, but the overall trend toward higher BH masses persists.  

This result is in conflict with theoretical evolutionary models of the formation of BH XRBs, which predict a continuous distribution of BH masses from neutron star masses up to 10--15~\msun\ \citep{fryer2001}.  Confirmation or elimination of the
low-mass BH ``gap'' would provide important constraints on the role of
massive star evolution, supernova energetics, and subsequent binary
evolution in BH formation \citep{fryer2001}. Further analysis of the mass distribution of BHs in compact binary systems is currently hampered by the generally poor precision of existing mass estimates (e.g., see Table 4.2 in McClintock \& Remillard 2005).  In this manuscript, we address this limitation by analyzing near-infrared (NIR) spectra to obtain a precise BH mass for one XRB, \vmon. 

\vmon\ was discovered in 1975 when it erupted in an X-ray nova
\citep{elvis1975}.  After its return to quiescence, \vmon\ was revealed as
an interacting binary system with a K star donating mass via an accretion
disk to a compact object \citep{oke1977,mcclintock1983}. Later, the
orbital period of the binary and the radial velocity amplitude of the donor
star were measured and yielded a mass function, $f(M) = 3.17$~\msun, which
established \vmon\ as a likely BH XRB \citep{mcclintock1986}.  Further
observations established the binary mass ratio and determined the masses of the
stars to within one unknown, the binary orbital inclination: $M_{1} =
(3.09\pm0.09)\sin^{3}i$~\msun\ and $M_{2} = (0.21\pm0.09)\sin^{3}i$~\msun,
where $M_{1}$ and $M_{2}$ are the BH and donor star masses, respectively
\citep{marsh1994, orosz1994}.  Several groups have determined values for the inclination, with the numbers ranging from $38\arcdeg \leq i \leq 75\arcdeg$.  As a result, estimates of the BH mass in \vmon\ vary from 3.3 to 13.6~\msun \citep{haswell1993,shahbaz1994,froning2001,gelino2001}.

The broad range of derived inclinations and BH masses for
\vmon\ result from long-term variability in the system and from
uncertain determinations of the amount of veiling, or dilution, by sources
other than the donor star.  The inclination is determined by modeling the amplitude of the ellipsoidal variations in the donor star light curve, so both of these effects will alter the derived BH mass.   In particular, an additional source will dilute the amplitude of the ellipsoidal variation, leading to an underestimate of the inclination and a corresponding overestimate of the BH mass if not taken into account. 

The best way to determine the true donor star contribution in \vmon\ is to model its
spectrum, particularly in the NIR, where the late type donor star is expected to dominate. \citet{shahbaz1999} modeled a low S/N, K-band spectrum of \vmon, from which they concluded that the accretion disk contributes at most 27\% of the continuum in the NIR.  They fit only the $^{12}$CO bandhead at 2.29~$\mu$m, however, which is sensitive to both temperature and luminosity of the donor star and may be prone to metallicity effects in compact binary systems \citep{froning2001}.  \citet{harrison2007} also recently published a K-band spectrum of \vmon\ in which they confirmed that the $^{12}$CO absorption lines are anomalously weak. 
What is needed to settle the debate over the contribution of the accretion disk to the NIR spectrum of \vmon\ are higher S/N, broadband spectra. To this end, we have obtained and present here 0.8 -- 2.4~$\mu$m spectroscopy of \vmon\ obtained with SpeX at the NASA InfraRed Telescope Facility (IRTF).  This manuscript is organized as follows: \S~\ref{sec_observ} summarizes the data reduction and calibration steps; \S~\ref{sec_analysis} presents the data analysis and modeling of the donor star spectrum; and \S~\ref{sec_discussion}  gives discussion and conclusions.

\section{Observations and Data Reduction} \label{sec_observ}

We observed \vmon\ on 2004 January 8 -- 10 using SpeX on the IRTF
\citep{rayner2003}. The weather was clear with good seeing conditions
($\lesssim0.7\arcsec$) throughout the run.  We observed \vmon\
using the ShortXD mode, which has a cross-dispersed echelle configuration
and covers 0.8 -- 2.5~$\mu$m simultaneously in 6 orders.  All observations
were obtained through the $0\farcs5$ slit, resulting in a spectral
resolution in the center of each order of R = 1200  (250~km~s$^{-1}$).  A nearby A0V star was observed hourly to sample the atmospheric absorption spectrum.  We also observed several spectral type calibration stars in the same configuration used for \vmon\ (supplemented with similar data from a previous SpeX observing run).  The observations are summarized in Table~\ref{tab_obs}.

All data reduction, calibration, and spectral extraction steps were performed
using Spextool, an IDL-based package developed by the IRTF \citep{cushing2004,vacca2003}.  The calibration processing
steps included flat-fielding, sky subtraction, optimal spectral extraction,
and wavelength calibration.  Each \vmon\ exposure was extracted individually to
preserve the full 300 sec time resolution.  The spectra were corrected for telluric absorption and flux-calibrated using the A0V stellar spectra as described by
\citet{vacca2003}. Finally, the orders were merged for each exposure to
create single spectra covering the full wavelength range.  The S/N per resolution element was $\sim$4 in the individual spectra. The slit was kept aligned to the parallactic angle during data acquisition, so the relative fluxes along the full 0.8 -- 2.4~$\mu$m range are accurate to $\leq2$\% (where the upper limit is the uncertainty at 0.9~$\mu$m when guiding at 2.4~$\mu$m;  Cushing et al. 2004).  In addition, stable observing conditions resulted in absolute flux calibrations of comparable quality.  We have not quantified this number as our analysis does not depend on the absolute flux of the spectrum, but we note for completeness that a rough comparison of our mean in-band colors to those of \citet{froning2001} and \citet{gelino2001} yielded JHK colors within 0.1 mag of their results, well within the level of variability observed in \vmon\ over long time periods.

For the time-averaged spectrum of \vmon, we shifted each 300 sec spectrum
to the rest frame of the donor star before median combining, using the orbital ephemeris of \citet{mcclintock1986} and the donor star radial velocity amplitude from
\citet{marsh1994}. The error bars were determined by calculating the median absolute deviation of each pixel and then propagated through the smoothing and de-reddening steps. We did not correct for the effects of orbital smearing within an exposure time, but we note that this is a negligible effect ($\leq$30~km~s$^{-1}$) at our 250~km~s$^{-1}$ spectral resolution. The time-averaged spectrum is shown in
Figure~\ref{fig_a0620}.  The spectrum has been boxcar-smoothed by 3 pixels,
equivalent to one resolution element. Based on the scatter around linear fits to (relatively) line-free spectral regions, we find that the S/N in the time-averaged spectrum is $\sim$55 in H and K and $\sim$45 in J ($>1 \mu$m, $\sim$30 at shorter wavelengths). 

Also shown in Figure~\ref{fig_a0620} is the dereddened spectrum of \vmon, calculated assuming a reddening along the line of sight of E(B--V) = 0.39 \citep{wu1976}. Figures~\ref{fig_a0620j} --~\ref{fig_a0620k} show expanded views of the J,H, and K bands of the spectrum, with prominent spectral absorption and emission features labeled.  Line identifications were made using multiple sources, including \citet{wallace2000}, \citet{meyer1998}, \citet{kleinmann1986}, \citet{harrison2004}, and the Atomic Line List\footnote{http://www.pa.uky.edu/\~peter/atomic/}.

\section{Analysis} \label{sec_analysis}

The broadband NIR spectrum of \vmon\ shown in Figures~\ref{fig_a0620} --~\ref{fig_a0620k}  is characterized by a blue continuum on which are superposed broad (full width at zero intensities $\geq$4000~km~s$^{-1}$) emission lines of \ion{H}{1} and \ion{He}{2} and narrow (full width at half minima $\simeq$ 250 -- 400~km~s$^{-1}$) absorption lines of neutral metals, including transitions of \ion{Na}{1}, \ion{Mg}{1}, \ion{Al}{1}, \ion{Si}{1}, \ion{K}{1}, \ion{Ca}{1}, \ion{Ti}{1}, and \ion{Fe}{1}. The emission lines are believed to originate in the accretion disk, while the absorption features originate in the photosphere of the donor star.  The absorption spectrum is similar to that of a K star, with previous estimates of the spectral type ranging from K3V to K7V \citep{oke1977,gonzalez2004}.  

In Figure~\ref{fig_cmp}, we show the dereddened spectrum of \vmon\ compared to that of a field K5V star.  The K star has been normalized to \vmon\ near 2.29 $\mu$m, just blueward of the $^{12}$CO (2,0) bandhead.  The comparison immediately shows that a K5V (or hotter) star cannot be the only flux source in the NIR.  If the K5 star is scaled to the flux of \vmon\ in the K band, it exceeds the dereddened flux by $>$20\% at bluer wavelengths.  Adopting the K4V and K3V spectral types of \citet{gelino2001} and \citet{gonzalez2004} results in an even larger disparity between the expected and observed J- and H-band fluxes.  Even a spectral type as late at K7V exceeds the observed flux in \vmon\ by up to 10\% over most of the J and H bands when normalized to 100\% contribution in K. 

Modest increases in the assumed reddening along the line of sight do not reconcile the spectrum of \vmon\ with that of a K5V star.  Adopting a reddening of E(B--V) = 0.45 brings the 0.8~$\mu$m fluxes into agreement, but the template spectrum is still brighter than the spectrum of \vmon\ at longer wavelengths, including most of the J and H bands.  Because the relative reddening values between H and K are small, the reddening must be increased to E(B--V) $>$ 1.0 to bring the spectrum of the normalized K5V template below the spectrum of \vmon\ at all NIR wavelengths.  It is extremely difficult  to reconcile a reddening this high with the observed depth of the interstellar absorption feature at 2200~\AA\ in the spectrum of \vmon\ \citep{wu1976}. As a result, the fundamental conclusion remains:  if the donor star is the sole source of NIR emission in \vmon, its spectral type must be later than that of a K7V.  Otherwise, some level of dilution must be present.   

The absorption spectrum of \vmon\ resembles that of the K5V template, but there is at least one important difference between them:  the CO molecular absorption features in \vmon\ are significantly weaker relative to the metal lines than in the template spectrum.  This difference can affect determinations of the donor star contribution to the NIR spectrum. For example, the dilution analysis performed  by \citet{shahbaz1999} on the K-band spectrum of \vmon\ is unlikely to be a valid determination of the contribution of the donor star to the NIR spectrum, since they relied entirely on the relative strength of the $^{12}$CO 2.29~$\mu$m feature.  The weakness of the CO lines also suggests that other anomalies may exist in the spectrum, necessitating that its decomposition be undertaken over a wide wavelength range and using multiple line species and features.  Accordingly, we compare the line equivalent widths and line ratios in \vmon\ with those of field star populations and model the spectrum using both spectral type standards and synthetic spectra.

\subsection{Classification Based on Spectral Indices}

Studies using equivalent widths (EWs) and EW ratios to determine the
spectral type of a star or stellar population have been pursued by several
groups (e.g., Origlia, Moorwood, \& Oliva 1993; Ali et al.\ 1995;
F\"{o}rster Schreiber 2000; and sources therein). Of particular interest to
us is the work of F\"{o}rster Schreiber (2000; hereafter FS), who examined
H and K-band absorption lines to find temperature and luminosity indices and indices sensitive to dilution of the stellar spectrum by other sources.  FS was primarily interested in spectral trends in giant and supergiant stars as the dominant stellar source in extragalactic NIR spectra, but his analysis includes some dwarf stars as well.

For comparison, we calculated the EWs of several prominent stellar absorption
lines in the time-averaged spectrum of \vmon\ to compare to the spectral
indices in FS.  Table~\ref{tab_ews} gives the measured EWs.  The lines were
chosen to correspond to those in Table~3 of FS and were calculated using
the same continuum normalization and integration limits.  Where applicable,
we also applied the EW correction for spectral resolution from Equations 2
-- 4 of FS.  The error bars on the EWs are the standard deviation of the mean for several measurements of each line with variable estimates of the continuum placement.

We first compared our EWs to the spectral indices given in
Figure 5 of FS, which presents temperature and luminosity indicators for stars with solar or near-solar abundances. With the exception of \ion{Si}{1} $\lambda$1.59~$\mu$m and \ion{Mg}{1} $\lambda$2.28~$\mu$m, all of the lines under analysis show a strong trend of increasing EWs with decreasing stellar temperature.  The EWs in \vmon\ are on the low side of the distributions for these lines, indicating stellar temperatures of $\geq$5000~K. 

We also compared our EWs with those presented in \citet{ali1995}, who concentrated on temperature indices for dwarf stars.  Note that the EWs in Table~\ref{tab_ews} used to compare to the \citet{ali1995} indices are larger than those used for the FS indices because Ali et al.\ used a wider wavelength interval for their measurements, which we mirrored. Using their EW-temperature relationships, we obtain a temperature of 4600$\pm$300~K from \ion{Ca}{1} and 5000$\pm$450~K from \ion{Na}{1}.   Therefore, if uncorrected for dilution, EWs in \vmon\ point to a donor star of roughly type K3V star or earlier.  However, stars of spectral type K7V or earlier exceed the observed spectrum of \vmon\ in J and H when zero dilution is assumed in K. Therefore, we conclude that a diluting continuum source must be present in the NIR spectrum of \vmon.  

The amount of dilution of the stellar spectrum by another source is
determined in FS by comparing the line ratios of adjacent atomic and
molecular features (their Table 8). Unfortunately, their line ratios use the H and K band CO molecular absorption features, which we have already seen are not normal in \vmon.  Indeed, a comparison of the $^{12}$CO 1.62~$\mu$m and 2.29~$\mu$m features in \vmon\ gives a result so disproportionately strong in the 1.62~$\mu$m line that the ratio doesn't even appear on the FS spectral index plot. Similarly, a comparison of the 2.29~$\mu$m feature to the nearby \ion{Na}{1} and \ion{Ca}{1} EWs indicates that the CO feature is weaker in \vmon\ than in any of the dwarf stars analyzed by FS.  These results indicate that the CO features cannot be used to estimate the non-stellar dilution component \vmon. 

\subsection{Fitting Spectral Type Standard Stars to the Spectrum of \vmon} \label{sec_template}

In an effort to quantify the contribution of the donor star to the NIR spectrum of \vmon\ and its dilution by other sources, we compared its spectrum to those of K3V, K5V, and K7V spectral type standard stars.  The standard stars (listed in Table~\ref{tab_obs}) were observed with the same instrument configuration and calibrated using the same procedures as for the \vmon\ observations.  Before comparing the \vmon\ and template spectra, both were boxcar-smoothed over 3 pixels (one resolution element).  The spectrum of \vmon\ was also dereddened with E(B--V) = 0.39 (using the reddening curve of Cardelli, Clayton, \& Mathis (1989) and the standard star spectra were convolved with a Gaussian of 83~km~s$^{-1}$ FWHM to mimic the rotational broadening of the donor star in \vmon\ \citep{marsh1994}.  Note, however, that the rotational broadening is smaller than the 250~km~s$^{-1}$ resolution of the spectra and has a minimal effect on the results.

We wrote an IDL program to fit scaled template spectra to the spectrum of \vmon\ using the following steps.  First, we selected a small wavelength range (typically, 0.1~$\mu$m or smaller) and fit a spline function to the continuum.  The continuum points were selected by eye.  After normalizing both the spectrum of \vmon\ and that of the template star and placing both spectra on a common, linear dispersion, we multiplied the template spectrum by a fraction, $f$, which represents the donor star contribution to the spectrum of \vmon, and subtracted the scaled donor star spectrum from that of \vmon.  We varied $f$ from 0 to 1 in increments of 0.01 to find the fraction that minimized the rms of the residual in each waveband.  Finally, we repeated this analysis over several spectral lines and groups of lines over the full NIR spectral range.  The fit regions we examined are given in Table~\ref{tab_ranges}.  The resulting best values of $f$ and the rms for each fit region and template spectrum is given in Table~\ref{tab_template}. 

There are few absorption lines in the J band that are both relatively strong and uncontaminated by emission lines in the \vmon\ spectrum, so our fits were restricted to the portion of the long J band between P$\gamma$ and P$\beta$ (1.10 -- 1.26~$\mu$m).  This region contains a blend of singly-ionized atomic species, including transitions of \ion{Mg}{1}, \ion{Fe}{1}, and \ion{Si}{1}.  The best fit fractions range from 0.78 to 1.0.  There is a disparity between the strongest line in this range, \ion{Mg}{1} $\lambda$1.183~$\mu$m and the other lines in this band: the \ion{Mg}{1} line is best fit at $f \sim$0.9, but the other lines are weaker in the template than in \vmon\ even at $f$ = 1.

The situation is less ambiguous in the H band.  The best fit to the full H band spectrum using the K5V standard star is shown in Figure~\ref{fig_hfit}.   The fit has  $f = 0.76$ and an rms of 0.016.  Although the $\chi^{2}$ statistics are relatively poor ($\chi^{2}_{\nu}$ = 13.5), there is a good qualitative correspondence between the morphologies of the observed and template spectra. Similar results are obtained with the K3V and K7V templates. Many of the stronger transitions (predominately \ion{Mg}{1} and \ion{Si}{1} lines) are too weak in the $f$ = 0.76 template, however.  If we restrict the fits to narrower regions around these lines, we generally obtain higher $f$ and better fits (e.g., $\chi^{2}_{\nu}$ = 5.2 for the 1.48 -- 1.52~$\mu$m region fit by the K5V template).  

The large $\chi^{2}_{\nu}$ values for our fits indicate that our error bars are undersized relative to the true uncertainty in the fits.  This is unsurprising, given the systematic uncertainties that affect modeling of NIR spectra in faint compact binaries, including the influence of sky background and telluric absorption correction, uncertain placement of the continuum level where no true continuum exists, and complex line blending wherein small temperature and/or abundance variations between the template and target stars can affect fit results.  As a result, we have chosen to determine the mean value and uncertainty in the H-band donor star contribution by using an average of fits to multiple lines and multiple templates over the full waveband, rather than thorough the use of $\chi^{2}$ statistics, as we believe the scatter between line fits provides a more rigorous sampling of uncertainties, particularly systematics. We determined the best representative value for the H band donor star contribution by averaging the best-fit $f$ values for the three narrowband H fit regions (1.48 -- 1.52~$\mu$m, 1.56--1.61~$\mu$m, and 1.70--1.72~$\mu$m) and the K5V and K7V templates. Averaging these values for the K5V and K7V template stars gives a donor star fraction $f$ = 0.82$\pm$0.02.  

In K, we fit three regions:  the spectrum shortward of \ion{He}{1} $\lambda$2.058~$\mu$m, which is dominated by \ion{Ca}{1} absorption lines; the region between \ion{He}{1} and B$\gamma$, which includes transitions of \ion{Mg}{1}, \ion{Al}{1}, and \ion{Si}{1}; and the region longward of B$\gamma$, which contains a rich blend of features, including transitions of \ion{Na}{1}, \ion{Ca}{1}, \ion{Fe}{1}, \ion{Ti}{1}, \ion{Mg}{1}, and CO.  Note that we do not have any fit results for the K3V template to the K-band spectrum, because the absorption lines in the K3V template are too weak relative to those of \vmon, even at $f$=1.  We have already shown from the comparison of the SEDs of \vmon\ and a K3V star that a star this hot cannot be the sole emission source (i.e., have $f = 1$) in \vmon\ in K without exceeding the observed spectrum at shorter wavelengths.  Now we see that a K3V star also cannot be reconciled to the spectrum of \vmon\ by decreasing its fractional contribution, because its absorption lines are already too weak at $f = 1$ to match those observed in \vmon.  As a result, we restrict our fits in K to K5V and K7V templates.  

The \ion{Ca}{1} lines from 1.90 -- 2.02~$\mu$m could not be simultaneously fit by a single value for $f$.  The fit values given in Table~\ref{tab_template} appear to the eye to be overdiluted as a result of spurious features (residuals of the telluric absorption correction) driving the fits.  The strongest lines --- 1.978 and 1.987~$\mu$m --- are fit by $f \sim 0.5$, but at this value the other lines in this region are too weak.  This spectral region may be contaminated by B$\delta$ $\lambda1.945$~$\mu$m emission from the accretion disk. Problems also beset the spectral fits in the 2.07 -- 2.15~$\mu$m region.  The K5V and K7V stellar spectra have anomalous emission bumps at 2.14~$\mu$m that cause visibly overdiluted fits over the full wavelength range.  If the fits are restricted to 2.07 -- 2.13~$\mu$m, best fits are obtained for $f \sim 0.65$, while fits to the strongest feature alone, \ion{Al}{1} 2.117~$\mu$m, gives $f = 0.75$ for both the template fits.  

The final fit region was the long-K portion of the spectrum, 2.18 -- 2.42~$\mu$m, which includes numerous atomic species transitions and the CO bandheads.  Figure~\ref{fig_kfit} shows the best fits for the K5V template over the full fit region and when the fit is restricted to $\lambda < 2.28$~$\mu$m, excluding the region dominated by the CO lines.  When long K is fit in its entirety, the fits are driven to large dilutions of the donor star to match the weak CO absorption in the \vmon\ spectrum.  At these low donor fractions ($f = 0.45$ for the K5V template and $f = 0.37$ for the K7V template), the atomic lines are too weak in the template spectrum relative to those in \vmon. If the fit is restricted to 2.18 -- 2.28~$\mu$m, the donor fractions rise to $f = 0.81$ (K5V) and $f = 0.76$ (K7V).  At these values, the atomic absorption spectrum of \vmon\ is well fit although there remain discrepencies between template and target spectra, most notably in the red components of \ion{Na}{1} $\lambda$2.209 and $\lambda$2.339~$\mu$m and in the \ion{Fe}{1} lines from 2.226 -- 2.247~$\mu$m, all of which are too weak in the template relative to \vmon.  A single discrepancy in \ion{Fe}{1} may explain these deviations, as there are \ion{Fe}{1} transitions coincident with both of the "\ion{Na}{1}" lines (see Figure 6 of FS for an illustration of the complex line blending in this spectral region).

\subsection{Fitting Model Spectra to the Spectrum of \vmon} \label{sec_linbrod}

In addition to modeling \vmon\ with standard star spectra, we used the LinBrod program to generate synthetic spectra for a Roche lobe-filling star in a compact binary with the geometry of \vmon\ \citep{bitner2006}.  We adopted the following parameter values for the models: $P_{orb} = 0.323$~d, $q = 0.067$, $i = 41\arcdeg$, and $K_{2} = 433$~km~s$^{-1}$ \citep{mcclintock1986, marsh1994,gelino2001}. We created phase-resolved spectra for donor star temperatures of T = 4000, 4250, 4500, 4750, and 5000~K.  Finally, we also created models at each temperature in which the carbon abundance in the star was reduced to [C/H] = -0.5, -1.0, -1.5, and -2.0. To compare to the observed spectrum of \vmon, we averaged the model spectra over the binary orbit after removing the donor star orbital motion and smoothed the spectra to the observed spectral resolution.  

Because the SED and spectral type standard star fits to the \vmon\ spectrum point to a donor star spectral type later than K3V, or $T < 5000$~K, we concentrated on the T = 4000, 4250, and 4500~K models.  In an effort to characterize the carbon depletion required to match the observed CO line depths, we also focused on the long-K portion of the spectrum (2.18 -- 2.42~$\mu$m) . Table~\ref{tab_linbrod} gives fit results for the LinBrod model fits.  We first fit the solar abundance models and then repeated the fits for the carbon-depleted spectra. Figure~\ref{fig_kCO} shows the normalized long K spectrum of \vmon\ with the solar abundance, T = 4000~K model and with the T = 4000~K, [C/H] = -1.5 spectrum.  The model spectra has been scaled by $f = 0.77$, the best fit donor fraction for the 2.18 -- 2.28~$\mu$m region.  The [C/H] = -1.5 models provided the best fit to the spectrum of \vmon\ for all three of the donor temperatures examined ($\chi^{2}_{\nu} = 4.0$ for the T=4000~K, [C/H] = -1.5 model fit to 2.28 -- 2.39~$\mu$m, versus $\chi^{2}_{\nu} = 5.9$ and $\chi^{2}_{\nu} = 5.2$  for the [C/H] = -1.0 and -2.0 models, respectively).  The -0.5 and -1.0 models had CO lines that were still too strong relative to those of \vmon, while the CO features were virtually absent in the -2.0 models and too weak to match the observed spectrum.  

\section{Discussion and Conclusions} \label{sec_discussion}

\subsection{The Donor Star in A0620--00}

\subsubsection{The Donor Star Spectral Type and Fractional Contribution to the NIR Spectrum}

Our analysis of the NIR spectrum of \vmon\ has demonstrated three principal results:  1) the donor star is not the only NIR flux source, with 18$\pm$2\% of the H-band flux originating in another component of the binary; 2) the donor star must be later than a K3V spectral type; and 3) the CO absorption lines are anomalously weak, requiring a carbon abundance of [C/H] = -1.5 in the donor star to match the observed line depths.

A comparison of the broadband NIR SED of \vmon\ with those of spectral type standard star spectra shows that the donor star cannot be the sole source of NIR flux.  If the standard star spectra are normalized to the dereddened spectrum of \vmon\ in the K-band, they exceed the observed flux in the J and H wavebands.  This result is true for all standard star spectra earlier than M0V. The discrepancy cannot be reconciled by changes in the differential reddening correction because the relative reddening correction between K and J and H is too small.  Additionally, the K-band absorption lines in the K3V standard star are too weak to match the line depths seen in the \vmon\ spectrum even at a 100\% donor star contribution.  Decreasing the donor star contribution to $f < 1$ will only make the template lines weaker, so a K3V spectral type for the donor star is ruled out.  

The most precise measurement of the donor star temperature is by \citet{gonzalez2004}, who fit synthetic spectra created from model atmospheres to the visible spectrum of \vmon.  They found that $T_{2} = 4900\pm100$~K, which corresponds to a K2/K3V spectral type.  This result is not in agreement with our requirement that the donor star be later than K3V.  There are reasons to believe that the \citet{gonzalez2004} results may be unreliable, however.  To determine the stellar parameters they used 24 \ion{Fe}{1} lines, to which they fit models with five free parameters:  stellar temperature, gravity, and metallicity, as well as a normalization and slope to represent dilution from the accretion disk.  The use of \ion{Fe}{1} lines alone to constrain all stellar parameters (plus the disk contribution) is uncommon practice for stellar modeling, which typically relies on independent determinations of the stellar temperature, as well as on both \ion{Fe}{1} and \ion{Fe}{2} transitions.  \citet{gonzalez2004} also determined that the abundances of the elements they fit were slightly above solar values.  The adoption of a cooler donor star temperature, as required by our results, will reduce their derived abundances. 

We conclude that the spectral type of the donor star is most likely between K5V to K7V, but do not attempt to constrain the spectral type more precisely.  The rms values for the dilution fits to various regions of the \vmon\ spectrum are comparable for both spectral types and we cannot rely on the CO features to create more precise temperature indices.  We therefore averaged the dilution values from both spectral type fits for the three narrow regions in the H-band to derive our H-band donor star fraction:  $f = 0.82\pm0.02$, or an 82\% donor star contribution in H. The donor fraction in long K ($>2.2 \mu$m) is 81\% for a K5V template or 76\% for the K7V.  Figure~\ref{fig_adisk} shows the spectrum of \vmon\ with a K5V template star scaled to 82\% of the H-band flux and the \vmon\ spectrum after the contribution of the donor star has been subtracted.  While the spectrum of \vmon\ is dominated by the K type donor, there is a significant second component consisting of a blue continuum and strong \ion{H}{1} and \ion{He}{2} line emission. 

Our results of a K5V to K7V donor star spectral type and 18\% -- 24\% veiling in H and K do not agree with those of \citet{gelino2001}, who found that the NIR SED of \vmon\  matched that of a K4V (T = 4600~K) star with no dilution.  As pointed out by \citet{hynes2005}, however, there is a degeneracy between the spectral type of the star and the amount and distribution of a veiling spectrum in an XRB:  a cooler donor star plus a diluting component can result in the same SED as a hotter donor star with no contamination. Hynes et al.\ showed that even a 100~K overestimate of the temperature of the donor star could result in a factor of two underestimate of the veiling.  Because our data show, both in the line EWs and the NIR SED,  that a K type donor star cannot be the only NIR flux source, we conclude that \citet{gelino2001} overestimated the temperature of the donor star and consequently underestimated the spectral dilution in the NIR.

Our results also disagree with the recent work of \citet{harrison2007}, who argue by analogy with the IR spectrum of the cataclysmic variable SS~Cygni that \vmon\ has $<$4\% dilution in the K-band.  Their argument can be summarized as follows:  a) \vmon\ and SS~Cyg have similar K-band spectral slopes, binary properties, and quiescent mass accretion rates; b) SS~Cyg also has a mid-IR excess; c) the NIR and MIR SEDs of SS~Cyg can be well fit by a K4V stellar spectrum plus free-free emission; d) because \vmon\ and SS~Cyg are similar, application of the star plus wind model can be extended from the latter to the former to estimate a $\sim$4\% limit on the contamination level of the NIR spectrum of \vmon. 

We have several concerns about this line of reasoning.  First, as already discussed above, the similarity of a SED to that of a field star does not preclude contaminating emission from the disk.  The data shown in Harrison et al.\ reinforce this:  the JHK colors of SS~Cyg are consistent with that of an undiluted K4V star even as the ellipsoidal light curves show clear evidence of contamination. In fact, Harrison el al.\ point out that the K-band spectrum of \vmon\ has a different slope from those of the field stars, which provides fairly clear evidence of contamination. Second, it is not known whether free-free emission is the correct explanation for the excess MIR flux observed in SS~Cyg.  Another possibility discussed by \citet{dubus2004}, who originally published the IR SED shown by Harrison et al., is that circumbinary disk emission is responsible. Indeed, \citet{muno2006} have found a MIR component in \vmon, which they fit with a $\simeq$600~K blackbody component, consistent with emission from circumbinary dust rather than free-free emission in a wind.  The cool blackbody component found by \citet{muno2006} does not affect the NIR spectrum of \vmon, which suggests that the attempt to use MIR data from SS~Cyg to estimate the NIR contamination in \vmon\ is a red herring.  Finally, we reiterate what our simultaneous, JHK spectra of \vmon\ indicate directly about the donor star contamination in the system:  based on the shape and fluxes of the JHK continuum spectrum, the absolute EWs of the atomic absorption lines, and dilution analyses using both field stars and a Roche-lobe filling model spectrum, the donor star must be significantly diluted ($\sim$20\%), even in the K-band.

\subsubsection{Weak CO Features in the Donor Star Spectrum}

Although the atomic absorption line spectrum in \vmon\ is consistent with a K5V -- K7V spectral type, the molecular $^{12}$CO lines are significantly weaker in \vmon\ than in a field star, corresponding to CO line strengths normally seen in early G stars.  Using the LinBrod model spectra of a Roche lobe-filling star, we found that reducing the C abundance to [C/H] = -1.5 results in a better match to the CO lines in the spectrum than the [C/H] = 0, -1.0, and -2.0 models when the donor star fraction is fit to $f = 0.77$. 

The weakness of the $^{12}$CO lines in \vmon\ has already been noted by \citet{harrison2007}, who determine that the $^{12}$C abundance must be decreased 50\% to match the depth of the $^{12}$CO bandheads.  The 50\% reduction in $^{12}C$ abundance ([$^{12}$C/H] = -0.3) found by Harrison et al.\ is significantly smaller than the 97\% drop we claim.  Again, however, we have several concerns about the method by which Harrison et al. obtained their results. First, they started with the line list included in the spectral synthesis program SPECTRUM \citep{gray1994}, but when they were unable to match the spectra of  standard spectral type field stars using this line list (and Kurucz atmospheres) due to the presence of strong absorption features in the model but not the observed spectra, they abandoned it and constructed one consisting only of  \ion{Na}{1}, \ion{Mg}{1}, and CO transitions.  When they found that the lines in their new models were too weak to match those of field stars at the correct temperature, they globally increased the $\log(gf)$ values of every line in their new line list to match up with observations.  They then applied this revised model to the spectrum of \vmon, adjusting the C isotopic abundances until the best fit by eye was achieved.

We are not confident in the quantitative reliability of spectral model fits in which spectral lines have been dropped and oscillator strengths altered in order to achieve even a rough fit to a K5V field star.  We note in contrast that our LinBrod models give fits to the spectrum of \vmon\ of comparable quality of those using template field stars with no deletions or alterations to the spectral synthesis line lists required.  Our other concern with the Harrison et al.\ fits is the placement of the continuum and evaluation of the best model fit.  They determined the best-fit model by eye.  To our eyes, however, the fits they show in their Figure 3 do not properly take into account the actual data quality:  their pseudo-continuum levels appear too high and their CO line minima are too low in their preferred model.  A rough comparison of their observed spectrum to ours shows that the $^{12}$CO normalized line depths are similar in both data sets, suggesting that a statistical model fit to their spectrum would result in a $^{12}$C abundance similar to the one we obtain.

Finally, Harrison et al.\ contend that in addition to a low $^{12}$C abundance in \vmon, the $^{13}$C abundance is enhanced, such that $^{13}$C/$^{12}$C = 1.  They  base their identification of $^{13}$CO on the presence of a feature coinciding with the  $^{13}$CO(3,1) 2.374~$\mu$m bandhead.  However, they also note that the $^{13}$CO(2,0) 2.345~$\mu$m is not seen in their spectrum, despite being the stronger feature of the two.  Given the poor atmospheric conditions at the time their data was taken and the increasing amount of telluric H$_{2}$0 absorption at these wavelengths, we do not believe that their data provide a clear detection of $^{13}$CO and certainly not of $^{13}$C at equal abundance with $^{12}$C.  In our spectrum of \vmon, there is a feature at the location of the $^{13}$CO(2,0) 2.345~$\mu$m bandhead, but no feature coinciding with $^{13}$CO(3,1) 2.374~$\mu$m.  We have marked the locations of the first two $^{13}$CO bandheads in our Figure~\ref{fig_a0620k}.  \citet{dhillon2002} point out, however, that a \ion{Ti}{1} feature is coincident with the $^{13}$CO(2,0) 2.345~$\mu$m bandhead and is a much more likely explanation of the feature we see.   As a result, we do not believe that an unambiguous detection of $^{13}$CO can be made in our spectrum and we find no evidence of any enhancement of this species in \vmon.

Anomalously weak CO absorption features have been seen in other compact binary systems.  The NIR spectra of several cataclysmic variables show weak or absent $^{12}$CO absorption \citep{harrison2000,harrison2004}. In the dwarf nova U~Gem, the NIR CO absorption lines are significantly weaker than in the M3V standard star spectrum that provides a good match to the atomic lines.  Model fits to the FUV spectrum of the metal-enriched white dwarf in U~Gem show that the C abundance on the WD surface is [C/H] = -1.0,  while the N abundance is highly super-solar, [N/H] = 0.7 \citep{sion1998,long1999,froning2001b}. Similarly, in the UV spectrum of the BH XRB XTE J1118+480 in outburst, typically strong emission lines of \ion{C}{4} and \ion{O}{5} are undetectable, while the \ion{N}{5} $\lambda$1240~\AA\ appears enhanced \citep{haswell2002}.  The UV line ratios are inconsistent with photoionization models, leading Haswell et al.\ to conclude that the emission spectrum is indicative of the accretion of C-depleted material from the donor star. As a result, \vmon\ can be added to the ever-increasing list of cataclysmic variables and XRBs that show depleted C (and, in some cases, enhanced N), pointing to a common history of nuclear processing of C to N in compact binary systems. 

\subsection{The Mass of the Black Hole in A0620--00}

Based on previous work, the mass of the BH in \vmon\ is known to the value of one unknown, the binary inclination:  $M_{1} = (3.09\pm0.09)\sin^{-3}i$ \citep{marsh1994}. The orbital inclination can be obtained by modeling the light curve of the donor star.  The donor star fills its Roche lobe and is distorted in
shape, which leads to a double-humped ellipsoidal variation in the light curve, the amplitude of which is dependent on inclination.  If a source besides the donor star contributes to the light curve, the amplitude of the ellipsoidal variation will be diluted, leading to an underestimate of the inclination and a corresponding overestimate of the BH mass if the contaminating source is not taken into account. 

The most precise inclination results to date were reported by \citet{gelino2001}, who modeled JHK light curves of \vmon. Based on good agreement between the target SED and that of a dereddened K4 star, they concluded that the donor star is the only NIR continuum source in \vmon.  Using this assumption, they modeled the ellipsoidal light curve and found $i = 40\fdg75\pm3\arcdeg$ and
$M_{1} = 11.0\pm1.9$~\msun. As discussed above, however, the NIR SED alone is insufficient to resolve the degeneracy between donor star temperature and veiling by another flux source in the system.  Our results indicate that the donor star cannot be the only NIR flux source in \vmon\ and that consequently,  the Gelino et al.\ results overestimate the BH mass in \vmon.

We previously modeled the H-band light curve in \vmon\ with a donor star plus accretion disk model and determined $38\arcdeg \leq i \leq 75\arcdeg$, or $3.3 \leq M_{1} \leq 13.6$~\msun\ \citep{froning2001}.  The broad range of values was caused by a degeneracy between the inclination and the fractional contribution of the accretion disk to the H-band light. Table~5 of \citet{froning2001} gives the inclination in \vmon\ as a function of the fractional contribution of diluting sources in the H-band. In \S~\ref{sec_template} of this paper, we determined that the donor star contributes 82$\pm$2\% of the H-band flux in \vmon. This result, a diluting fraction of 18$\pm$2\%, combined with Table~5 in \citet{froning2001} gives a binary inclination for \vmon\ of $i = 43\pm1\arcdeg$.  

Based on this inclination, we obtain the mass of the BH accretor in \vmon:  $M_{1} = 9.7\pm$0.6~$M_{\odot}$.  This result is comparable to previous estimates of the BH mass in the literature.  \citet{shahbaz1994} found a BH mass of 10~M$_{\odot}$, while \citet{gelino2001} found $M_{1} = 11.0\pm$1.9~$M_{\odot}$.  The inclination we derive is within the error interval of Gelino et al.  The difference in BH masses between the two results comes from the slightly higher inclination we adopt as a result of our determination that the donor star cannot be the sole NIR emission source. 

The error bar on our derived BH mass represents the propagated statistical errors in the result.  A potential source of systematic uncertainty is our assumption that the atomic absorption spectrum of \vmon\ can be modeled by template spectra with solar abundances.  The fact that the C abundance in \vmon\ has to be decreased significantly to match the CO lines suggests the need for caution in this regard.  However, the relative line ratios of the atomic transitions largely agree with each other in the derived fractional donor star contributions.  These lines are also strong transitions that lie on the flat portion of the curve of growth and will not be sensitive to small abundance variations. Finally, we note that while \citet{gonzalez2004} derived slightly super-solar abundances for the metal lines in \vmon, they used a stellar temperature that is too hot to be consistent with the NIR SED. The adoption of a cooler temperature for the donor star will cause the metal line abundances required to fit the optical spectrum to decrease.

Another potential source of systematic error is the large time interval (8 years) between acquisition of the NIR light curve and the spectra.  Our analysis assumed that the donor star diluting fraction found by analyzing the spectra applies equally well to the light curve data.  This assumption is valid, we believe, because while \vmon\ is variable, its NIR colors typically don't vary by more than 0.2~mag, and six observations of the H-band light curve spaced over days to years had mean colors that agreed to within 0.04~mag \citep{froning2001,gelino2001}.  Our rough estimate of the absolute calibration of our time-averaged spectrum was also consistent with the previous measurements.  We can estimate the uncertainty in the non-donor star contribution by assuming that it could vary by $\pm$4\%, consistent with the previous measurements.  This would cause the donor star fraction to range from $79 \leq f \leq 86$, which results in $i = 43\pm1\arcdeg$, consistent with our statistical uncertainties.

\acknowledgments
We thank Nathaniel Cunningham for assistance in observing
\vmon, and the staff at the IRTF for their support. We also thank Chris Sneden and Niall Gaffney for their help in calculating the LinBrod donor star spectra and for useful discussions.

Facilities: \facility{IRTF(SpeX)}

\clearpage
\begin{deluxetable}{lcccc}
\tablecaption{SpeX Observations \label{tab_obs}}
\tablewidth{0pt}
\tablecolumns{5}
\tablehead{
\colhead{Object} & \colhead{Date (UT)} & \colhead{Instrument} & 
\colhead{T$_{exp}$ (min)} & $\Phi$\tablenotemark{a} }
\startdata
A0620-00 & 2004 Jan 8 & SpeX & 290 & 0.59 -- 1.47 \\
A0620-00 & 2004 Jan 9 & SpeX & 280 & 0.69 -- 1.49 \\
A0620-00 & 2004 Jan 10 & SpeX & 250 & 0.92 -- 1.61 \\
HD42606 (K2.5 III) & 2004 Jan 18 & SpeX & 1.2 & \nodata \\
HD 3765 (K2 V) & 2004 Jan 18 & SpeX & 2 & \nodata \\
HD16160 (K3 V) & 2004 Jan 19 & SpeX & 1.8 & \nodata \\
61 Cyg A (K5 V) & 2000 Sept 15 & SpeX & 0.5 & \nodata \\
61 Cyg B (K7 V) & 2000 Sept 15 & SpeX & 0.5 & \nodata \\
\enddata
\tablenotetext{a}{Orbital phase coverage of A0620--00 for each night's observations, based on the ephemeris of McClintock \& Remillard (1986).}
\end{deluxetable}
e
\clearpage
\begin{deluxetable}{lcc}
\tablecaption{Equivalent Widths of Selected Absorption Lines \label{tab_ews}}
\tablewidth{0pt}
\tablecolumns{3}
\tablehead{
\colhead{Feature} & \colhead{EW\tablenotemark{a} (\AA)} & \colhead{EW\tablenotemark{b} (\AA)} }
\startdata
\protect\ion{Si}{1} 1.5892 & 1.52$\pm$0.04  & \nodata \\
$^{12}$CO (6,3) 1.6187 & 0.86$\pm$0.07 & \nodata \\
\protect\ion{Na}{1} 2.2076 & 0.94$\pm$0.06 & 2.18$\pm$0.04 \\
\protect\ion{Fe}{1} 2.2263 & 0.36$\pm$0.05 & \nodata \\
\protect\ion{Fe}{1} 2.2387 & 0.40$\pm$ & \nodata \\
\protect\ion{Ca}{1} 2.2636 & 1.06$\pm$0.06 & 3.21$\pm$0.03 \\
\protect\ion{Mg}{1} 2.2814 & 0.43$\pm$0.04 & 0.68$\pm$0.06 \\
$^{12}$CO (2,0) 2.2935 & 0.38$\pm$0.06  & 2.26$\pm$0.08 \\
$^{12}$CO (3,1) 2.3227 & 0.22$\pm$0.02 & \nodata \\
$^{13}$CO (2,0) 2.3448 & 0.64$\pm$0.04 & \nodata \\
\enddata
\tablenotetext{a}{EWs calculated using the integration limits of F\"{o}rster Schreiber 2000.}
\tablenotetext{b}{EWs calculated using the integration limits of Ali et al.\ (1995).}
\end{deluxetable}

\clearpage
\begin{deluxetable}{lll}
\tablecaption{Wavelength Ranges for Spectral Fits \label{tab_ranges}}
\tablewidth{0pt}
\tablecolumns{3}
\tablehead{
\colhead{Waveband} & \colhead{$\lambda$ Range ($mu$m)} & {Description} }
\startdata
J & 1.10 -- 1.26 & Long J \\
 & 1.175 -- 1.215 & Blend incl. \protect\ion{Mg}{1}, \protect\ion{Fe}{1}, \protect\ion{Si}{1} \\
H & 1.4 -- 1.8 & Full H band \\
  & 1.48 -- 1.52 & \protect{Mg}{1} lines. \\
  & 1.56 -- 1.61 & Blend incl. \protect\ion{Mg}{1}, \protect\ion{Si}{1} \\
  & 1.70 -- 1.72 & \protect{Mg}{1} \\
K & 1.90 -- 2.02 & Several \protect\ion{Ca}{1} lines. \\
  & 2.07 -- 2.15 & Short K incl. \protect\ion{Mg}{1}, \protect\ion{Si}{1}, \protect\ion{Al}{1}. \\
  & 2.18 -- 2.42 & Long K incl. CO bandhead. \\
  & 2.18 -- 2.28 & No CO, lines incl. \protect\ion{Na}{1}, \protect\ion{Fe}{1}, \protect\ion{Ca}{1}, \protect\ion{Mg}{1}. \\
\enddata
\end{deluxetable} 

\clearpage
\begin{deluxetable}{lccc}
\tablecaption{Template Star Fits to A0620--00 Spectrum \label{tab_template}}
\tablewidth{0pt}
\tablecolumns{4}
\tablehead{
\colhead{Template} & \colhead{Wavelength Range} & \colhead{Template Fraction} & \colhead{rms}   \\
\colhead{Spectral Type} & \colhead{($\mu$m)} & \colhead{(f)} }
\startdata
K3V & 1.10--1.26 & 0.78 & 0.012 \\
& 1.175 -- 1.215 & 0.82 & 0.010 \\
& 1.4--1.8 & 0.76 & 0.016 \\
& 1.48 -- 1.52 & 0.71 & 0.011 \\
& 1.56 -- 1.61 & 0.81 & 0.011 \\
& 1.70 -- 1.72 &  0.91 & 0.009 \\
K5V & 1.10 -- 1.26 & 0.90 & 0.013 \\
 & 1.175 -- 1.215 & 1.0 & 0.010 \\
 & 1.4 -- 1.8 & 0.78 & 0.016  \\
 & 1.48 -- 1.52 & 0.82 & 0.010 \\
 & 1.56 -- 1.61 & 0.88 & 0.015 \\
 & 1.70 -- 1.72 & 0.78 & 0.010 \\
 & 1.90 -- 2.02 & 0.37 & 0.02 \\
 & 2.07 -- 2.15 & 0.52 & 0.01 \\
 & 2.18 -- 2.42 & 0.45 & 0.012  \\
 & 2.18 -- 2.28 & 0.81 & 0.009 \\
K7V & 1.10 -- 1.26 & 0.87 & 0.013 \\
 & 1.175 -- 1.215 & 0.99 & 0.011 \\
 & 1.4 -- 1.8 & 0.76 & 0.017 \\
 & 1.48 -- 1.52 & 0.82 & 0.010 \\
 & 1.56 -- 1.61 & 0.85 & 0.015 \\
 & 1.70 -- 1.72 & 0.77 & 0.009 \\
  & 1.90 -- 2.02 & 0.39\tablenotemark{a} & 0.021 \\
 & 2.07 -- 2.15 & \nodata\tablenotemark{b} & \nodata \\
 & 2.18 -- 2.42 & 0.37 & 0.014 \\
 & 2.18 -- 2.28 & 0.76 & 0.009 \\
\enddata
\tablenotetext{a}{Fits too diluted due to noise in spectra.}
\tablenotetext{b}{Fits compromised by spurious feature in template.}
\end{deluxetable}

\clearpage
\begin{deluxetable}{ccccc}
\tablecaption{LinBrod Fits to A0620--00 Spectrum \label{tab_linbrod}}
\tablewidth{0pt}
\tablecolumns{5}
\tablehead{
\colhead{Model Temperature} & \colhead{Model C Abundance} & \colhead{Wavelength Range} & \colhead{Template Fraction} & \colhead{rms} \\
\colhead{(K)} & \colhead{(log[C/H]/[C/H]$_{\odot}$)} & \colhead{($\mu$m)} & \colhead{(f)} }
\startdata
4000 & 0.0 & 2.18 -- 2.28 &  0.72 & 0.011 \\
4250 & 0.0 & 2.18 -- 2.28 & 0.76 & 0.011 \\
4500 & 0.0 & 2.18 -- 2.28 & 0.99 & 0.011 \\
\cutinhead{Full Long K Region Including CO Lines} 
4000 & 0.0 & 2.18 -- 2.38 & 0.14 & 0.015 \\
4000 & -0.5 & 2.18 -- 2.38 & 0.28 & 0.014 \\
4000 & -1.0 & 2.18 -- 2.38 & 0.54 & 0.012 \\
4000 & -1.5 & 2.18 -- 2.38 & 0.78 & 0.011 \\ 
4000 & -2.0 & 2.18 -- 2.38 & 0.74 & 0.012 \\
4250 & 0.0 & 2.18 -- 2.38 & 0.14 & 0.015 \\
4250 & -0.5 & 2.18 -- 2.38 & 0.28 & 0.014 \\
4250 & -1.0 & 2.18 -- 2.38 & 0.56 & 0.013 \\
4250 & -1.5 & 2.18 -- 2.38 & 0.82 & 0.012 \\
4250 & -2.0 & 2.18 -- 2.38 & 0.81 & 0.012 \\
4500 & 0.0 & 2.18 -- 2.38 & 0.17 & 0.015 \\
4500 & -0.5 & 2.18 -- 2.38 & 0.35 & 0.014 \\
4500 & -1.0 & 2.18 -- 2.38 & 0.71 & 0.013 \\
4500 & -1.5 & 2.18 -- 2.38 & 0.96 & 0.012 \\
4500 & -2.0 & 2.18 -- 2.38 & 0.93 & 0.013 \\
\cutinhead{Longward of the 2.29~$\mu$m Bandhead Only\tablenotemark{a}}
4000 & -0.5 & 2.28 -- 2.38 & 0.77 & 0.025 \\
4000 & -1.0 & 2.28 -- 2.38 & 0.77  & 0.014 \\
4000 & -1.5 & 2.28 -- 2.38 & 0.77 & 0.012 \\
4000 & -2.0 & 2.28 -- 2.38 & 0.77 & 0.013 \\
4250 & -0.5 & 2.28 -- 2.38 & 0.77 & 0.024 \\
4250 & -1.0 & 2.28 -- 2.38 & 0.77 & 0.014 \\
4250 & -1.5 & 2.28 -- 2.38 & 0.77 & 0.012 \\
4250 & -2.0 & 2.28 -- 2.38 & 0.77 & 0.013 \\
4500 & -0.5 & 2.28 -- 2.38 & 0.77 & 0.020 \\
4500 & -1.0 & 2.28 -- 2.38 & 0.77 & 0.014 \\
4500 & -1.5 & 2.28 -- 2.38 & 0.77 & 0.013 \\
4500 & -2.0 & 2.28 -- 2.38 & 0.77 & 0.014 \\
\enddata
\tablenotetext{a}{Donor star fraction fixed in these models to the best-fit value to the nearby atomic lines from the template star fits.}
\end{deluxetable}

\clearpage

\figcaption[a0620.ps]{The NIR spectrum of A0620--00, obtained in 2004 January. The solid line shows the time-averaged spectrum of A0620.  Individual exposures were shifted to remove the orbital motion of the donor star before averaging. The dotted line shows the spectrum after dereddening, assuming E(B--V) = 0.39 (Wu et al. 1976). \label{fig_a0620}}

\figcaption[a0620j.ps]{The spectrum of A0620-00 in J.  Prominent spectral
  features are labeled. An error bar representative of the statistical uncertainty per resolution element is plotted on the far right of the plot. Also shown at the bottom of the figure is the spectrum of HD45137, the A0V star used for telluric correction of the A0620--00 spectra.  The spectrum retains the throughput profile of each spectral order but is not shown with absolute flux calibration.  The \protect\ion{H}{1} lines intrinsic to the A0V spectrum have been fitted and removed using the xtellcor program developed by the IRTF. \label{fig_a0620j}}

\figcaption[a0620h.ps]{The spectrum of A0620-00 in H.  Prominent spectral
  features are labeled. A representative error bar for a resolution element is plotted on the far right. The telluric spectrum is also shown at the bottom of the figure. \label{fig_a0620h}}

\figcaption[a0620k.ps]{The spectrum of A0620-00 in K.
Prominent spectral features are labeled. A representative error bar for a resolution element is plotted on the far right. The telluric spectrum is also shown. \label{fig_a0620k}}

\figcaption[compare.ps]{Shown in black is the dereddened spectrum of A0620-00.  Shown in gray is the spectrum of  61 Cyg A, a K5V spectral type star. The spectrum of 61 Cyg A has been normalized to the flux of A0620 just blueward of the $^{12}$CO 2.29~$\mu$m bandhead.  \label{fig_cmp}}

\figcaption[hfit.ps]{The normalized H-band spectrum of A0620--00 with a scaled spectral type standard star fit.  The standard star, shown in red,  is 61 Cyg A, a K5V star.  It has been scaled by $f = 0.76$. \label{fig_hfit}}

\figcaption[kfit.ps]{The normalized K-band spectrum of A0620--00 with a scaled spectral type standard star fit.  The template star is 61 Cyg A, a K5V star.  The solid red spectrum shows the template scaled by $f = 0.45$, the best fit over the full 2.18 -- 2.42~$\mu$m range. The dashed red spectrum shows the template scaled by $f = 0.81$, the best fit to the 2.18 -- 2.28~$\mu$m region. \label{fig_kfit}}

\figcaption[klind.ps]{The normalized K-band spectrum of A0620--00 with scaled LinBrod T = 4000 K model spectra fits.  The solid red line shows the LinBrod model with [C/H] = -1.5, while the dashed red line shows the solar abundance model.  To avoid confusion, the latter is shown only for $\lambda > 2.288$~$\mu$m. The models are scaled by $f = 0.77$. \label{fig_kCO}}

\figcaption{The top panel shows the dereddened spectrum of \vmon\ in black and the spectrum of 61 Cyg A, a K5V star, in gray.   The spectrum of the K5V template has been scaled to 82\% of the \vmon\ flux at the center of the H band, 1.6~$\mu$m. The lower panel shows the NIR spectrum of the accretion disk in \vmon, created by subtracting the template spectrum from that of \vmon. \label{fig_adisk}}

\clearpage
\begin{figure}
\plotone{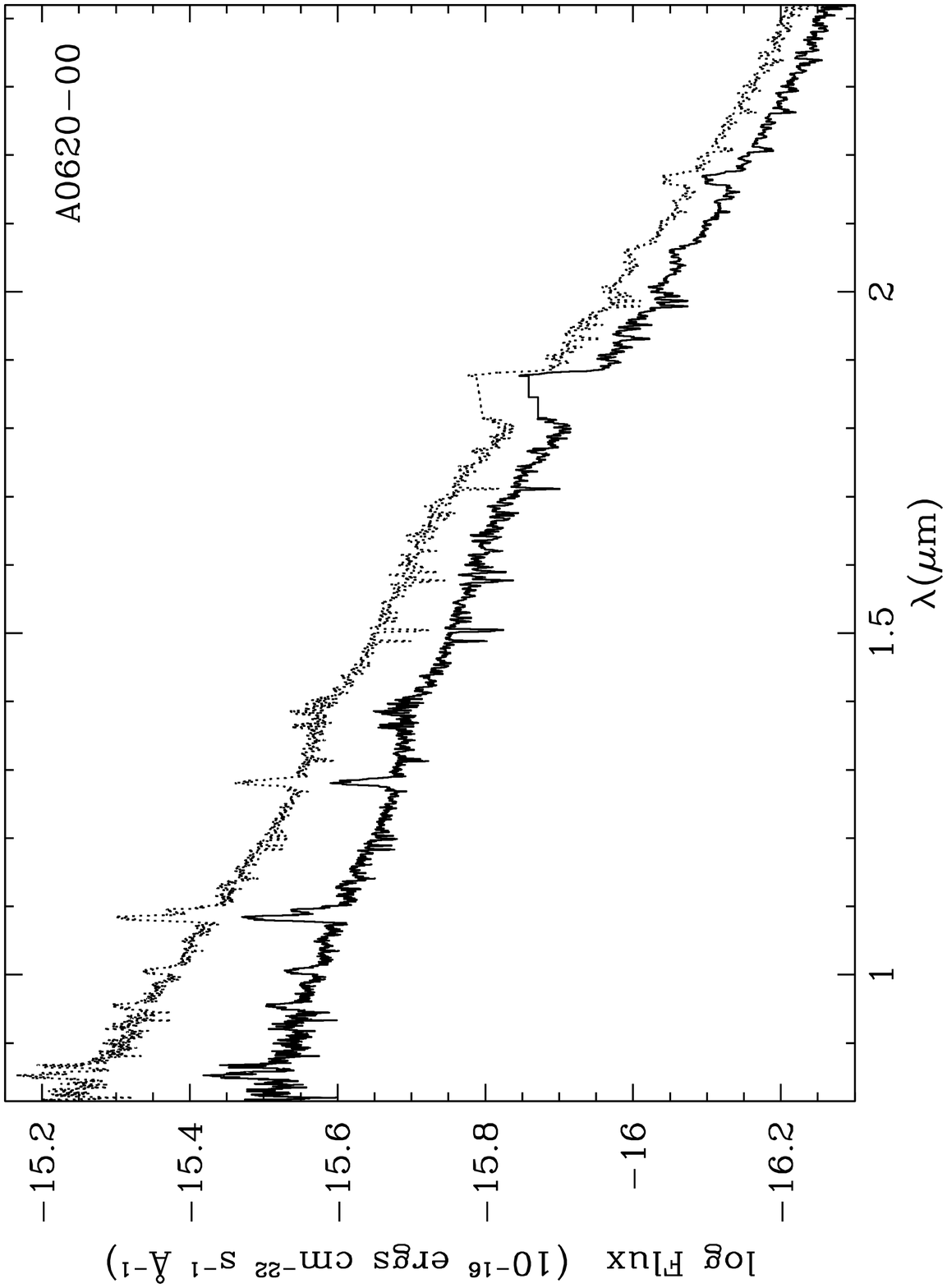}
\centerline{f1.eps}
\end{figure}
\clearpage

\begin{figure}
\plotone{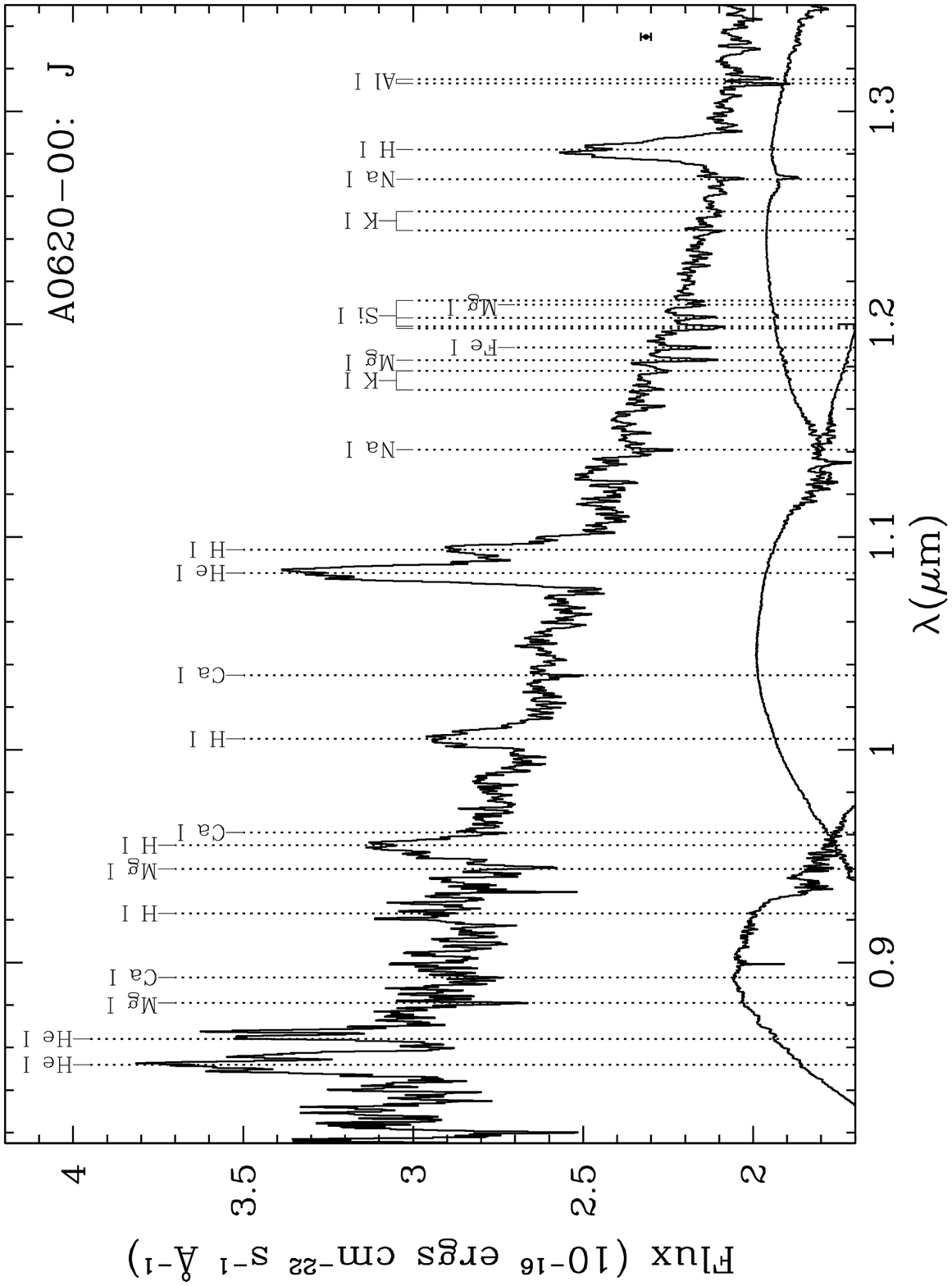}
\centerline{f2.eps}
\end{figure}
\clearpage

\begin{figure}
\plotone{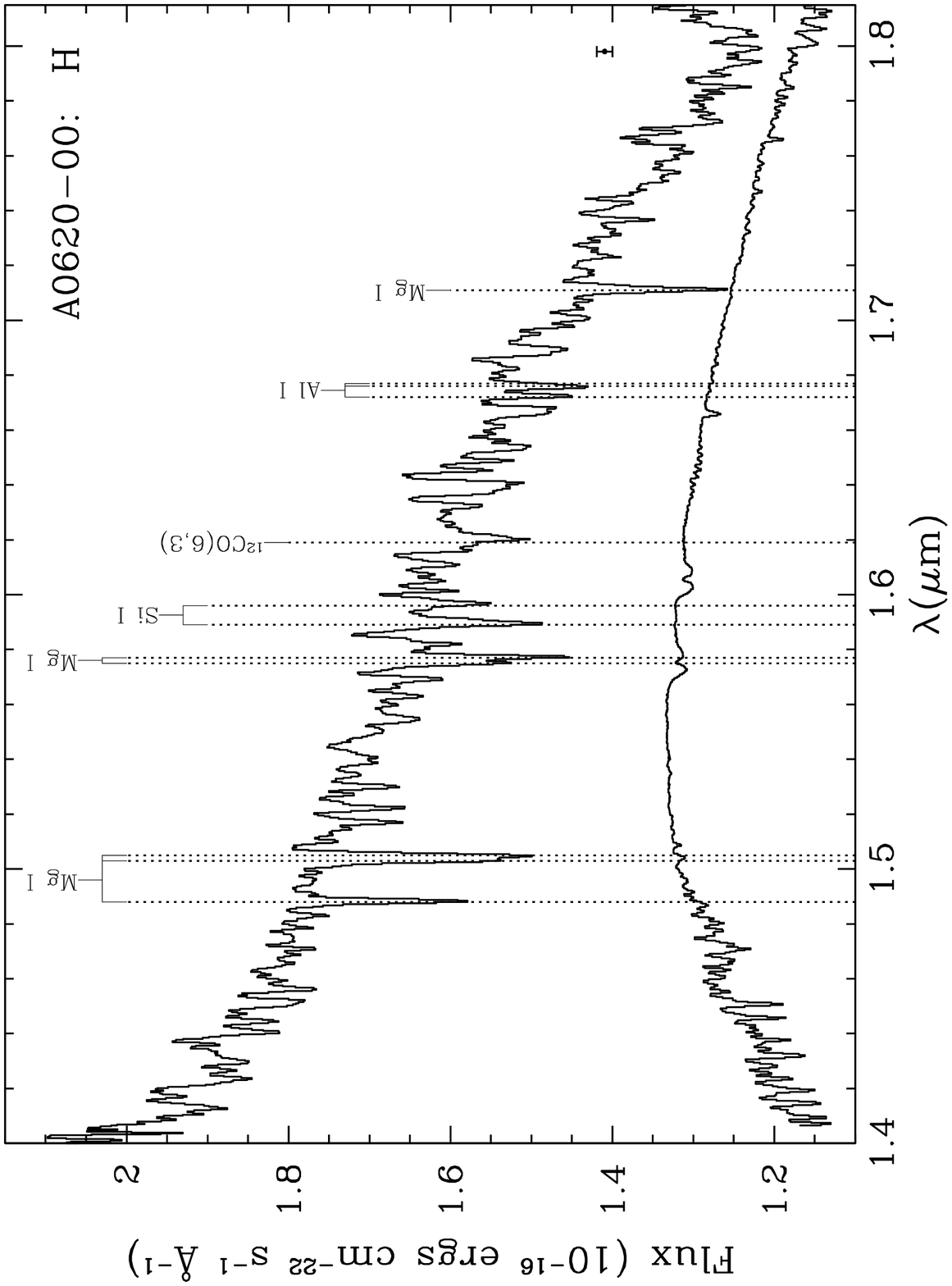}
\centerline{f3.eps}
\end{figure}
\clearpage

\begin{figure}
\plotone{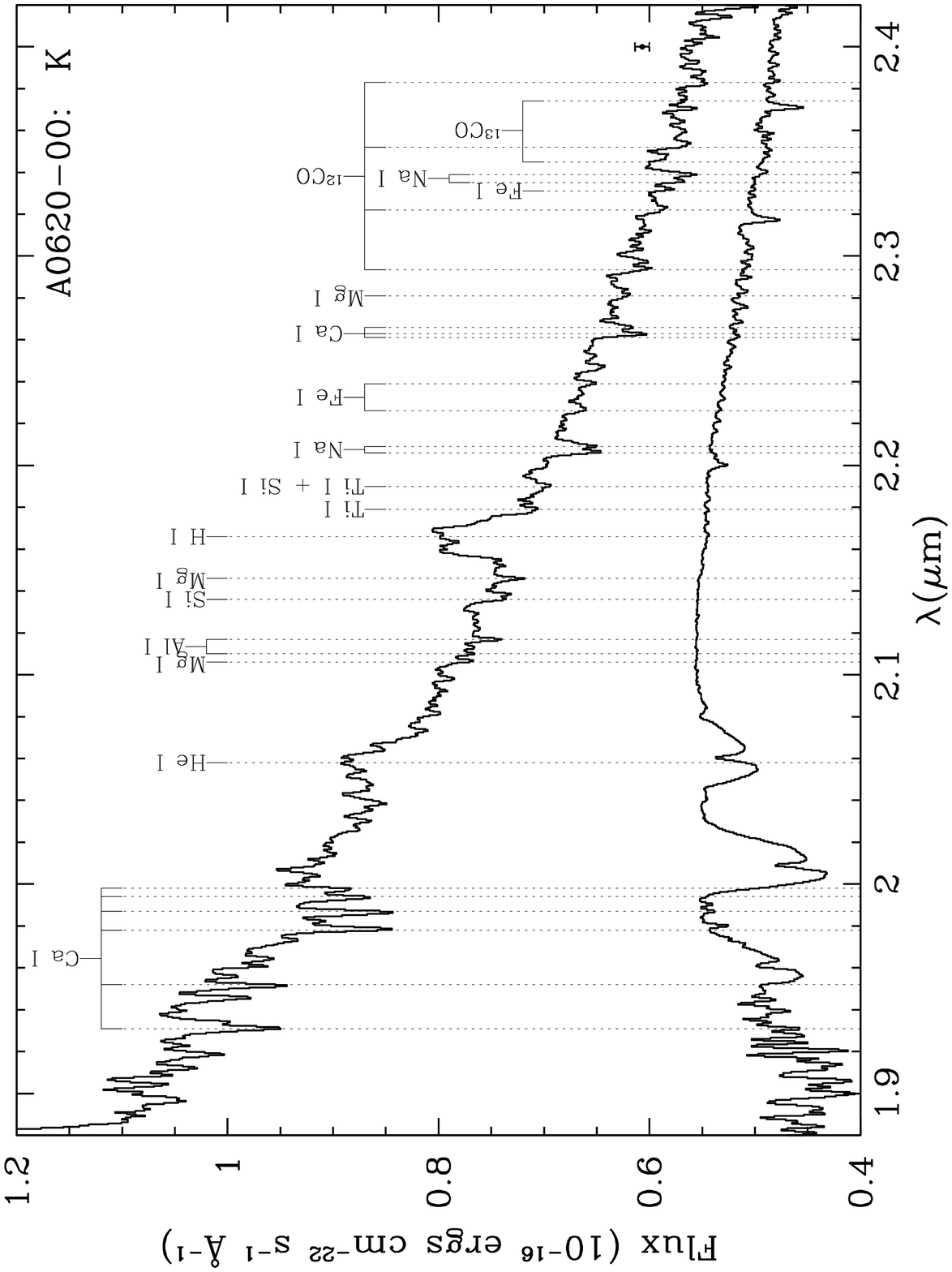}
\centerline{f4.eps}
\end{figure}
\clearpage

\begin{figure}
\plotone{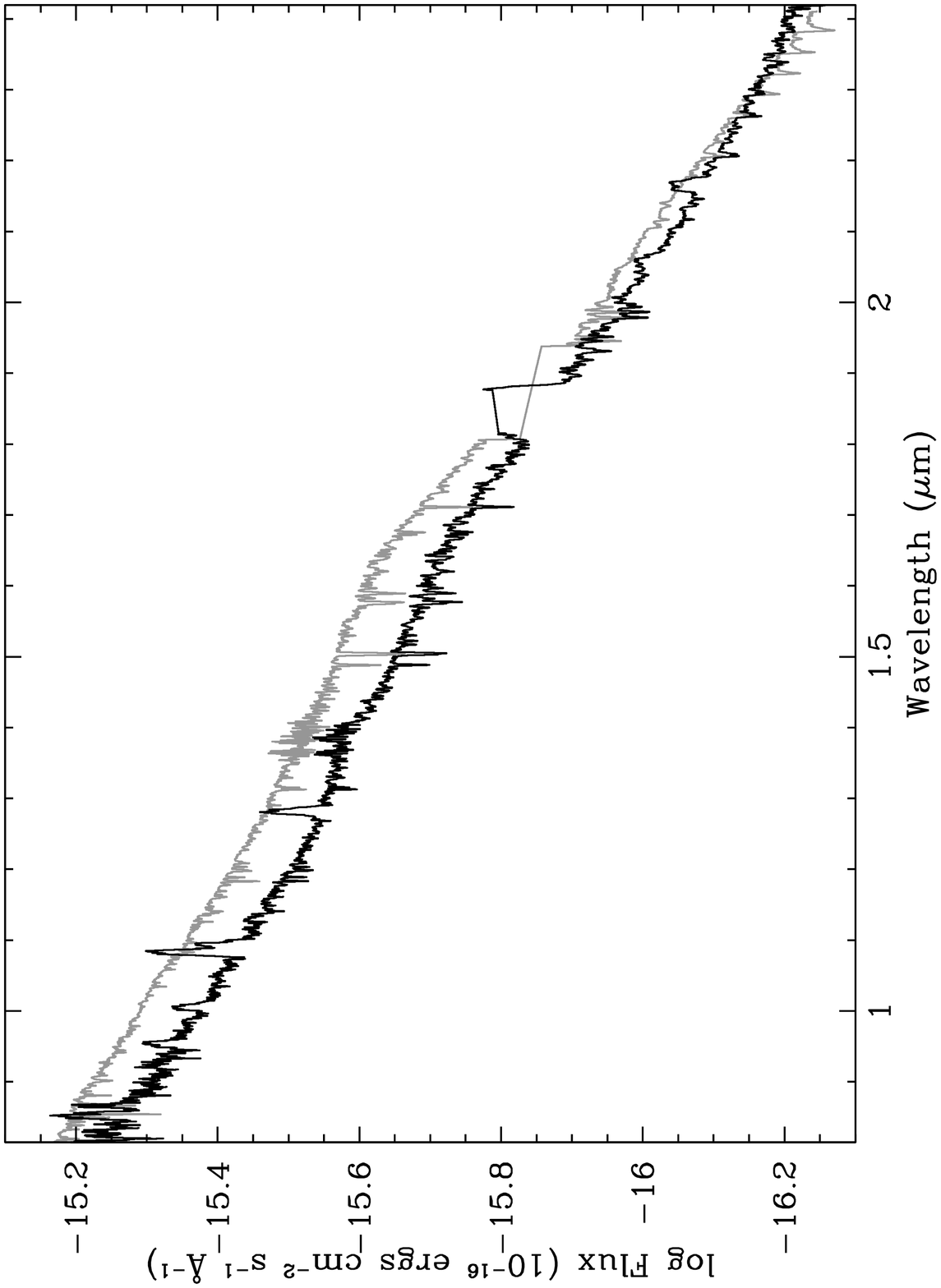}
\centerline{f5.eps}
\end{figure}
\clearpage

\begin{figure}
\plotone{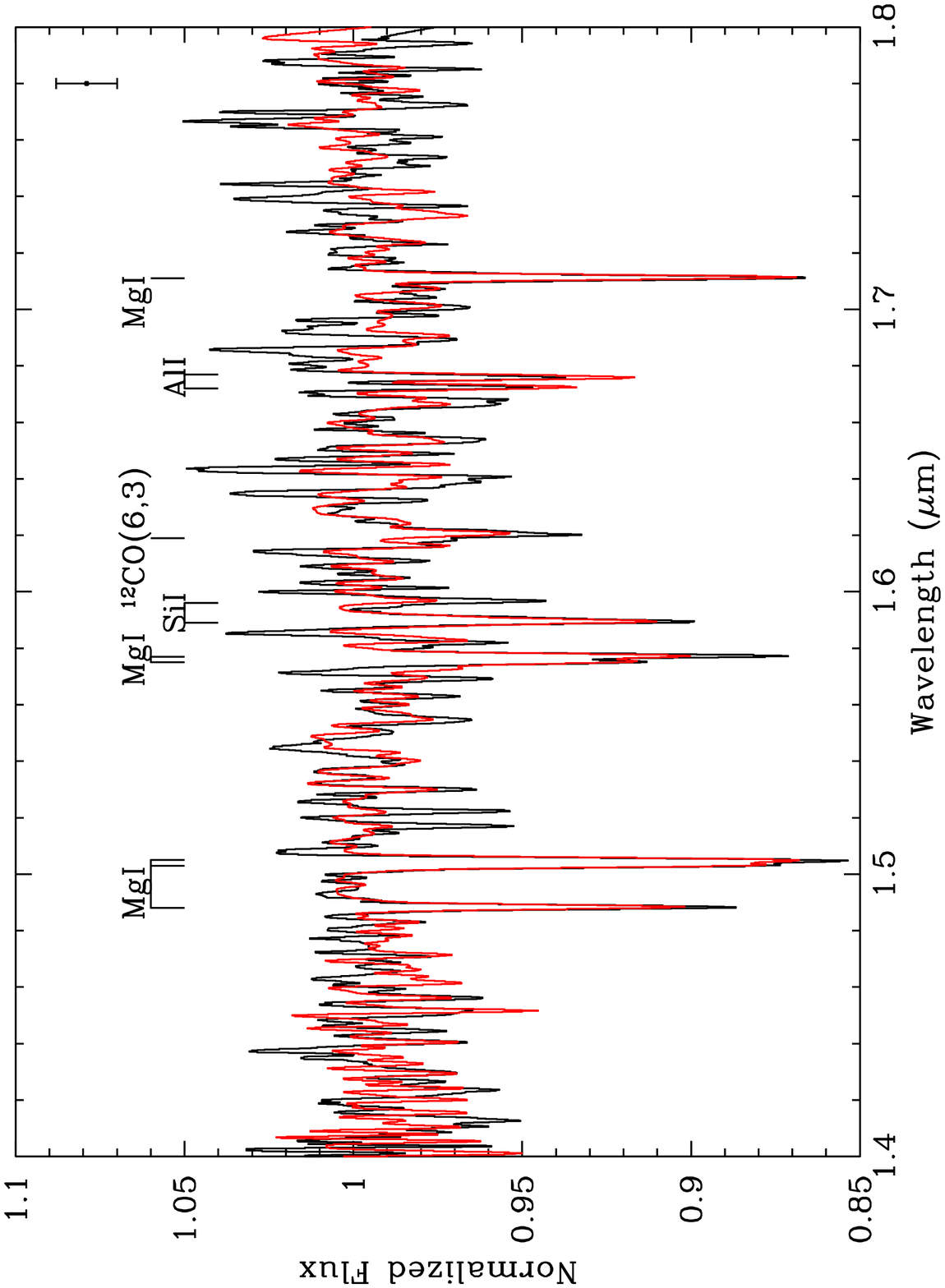}
\centerline{f6.eps}
\end{figure}
\clearpage

\begin{figure}
\plotone{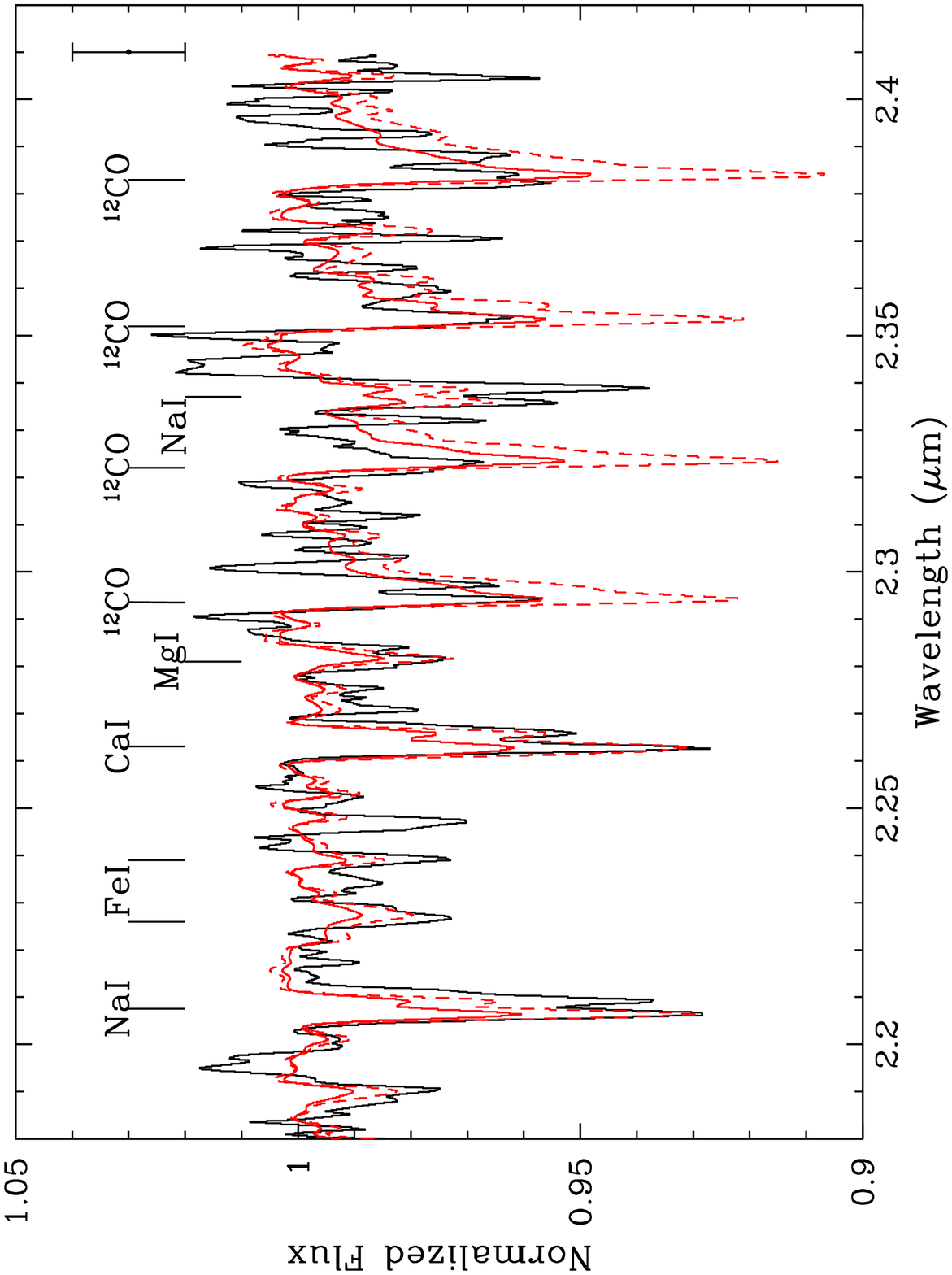}
\centerline{f7.eps}
\end{figure}
\clearpage

\begin{figure}
\plotone{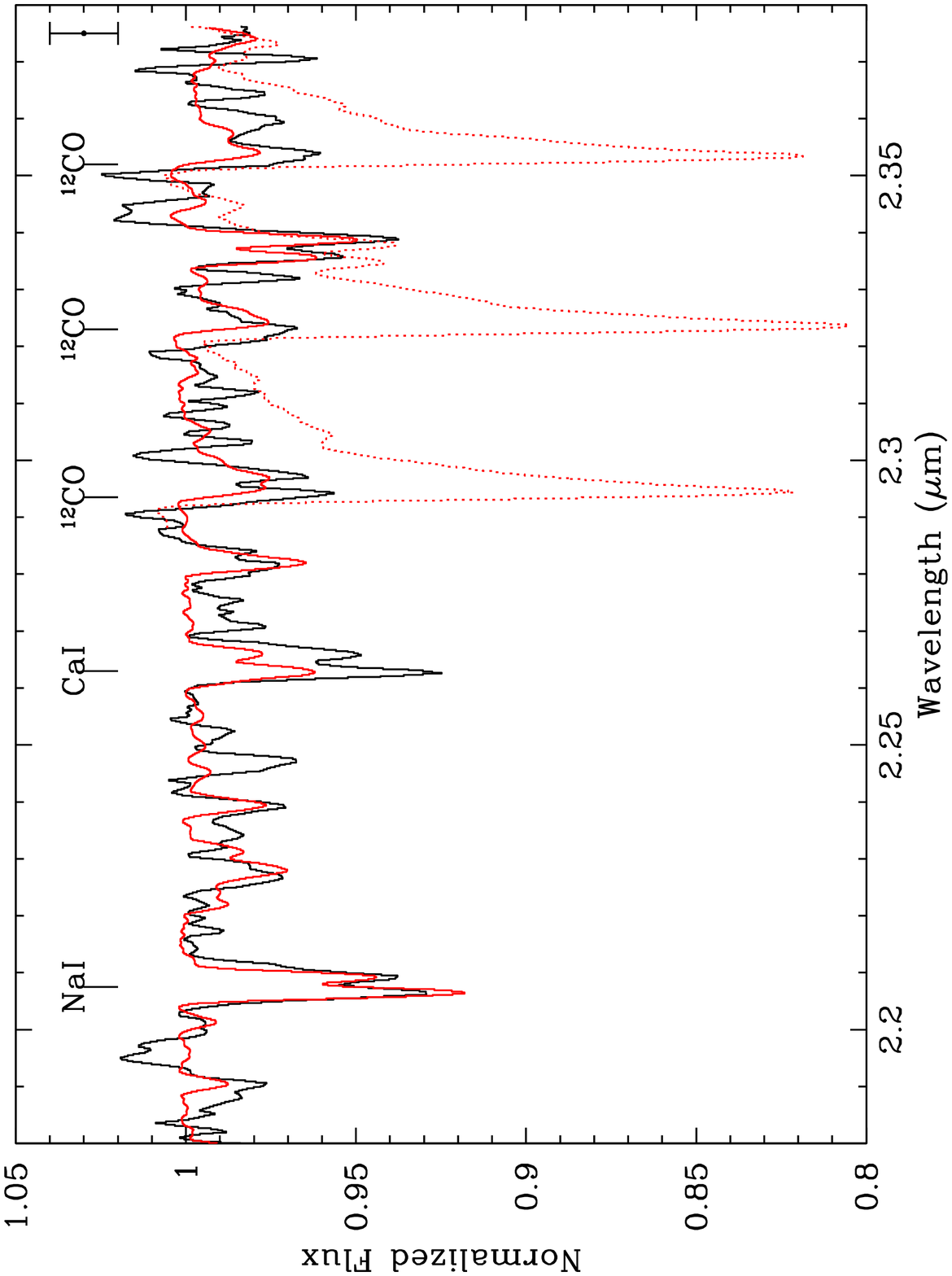}
\centerline{f8.eps}
\end{figure}
\clearpage

\begin{figure}
\plotone{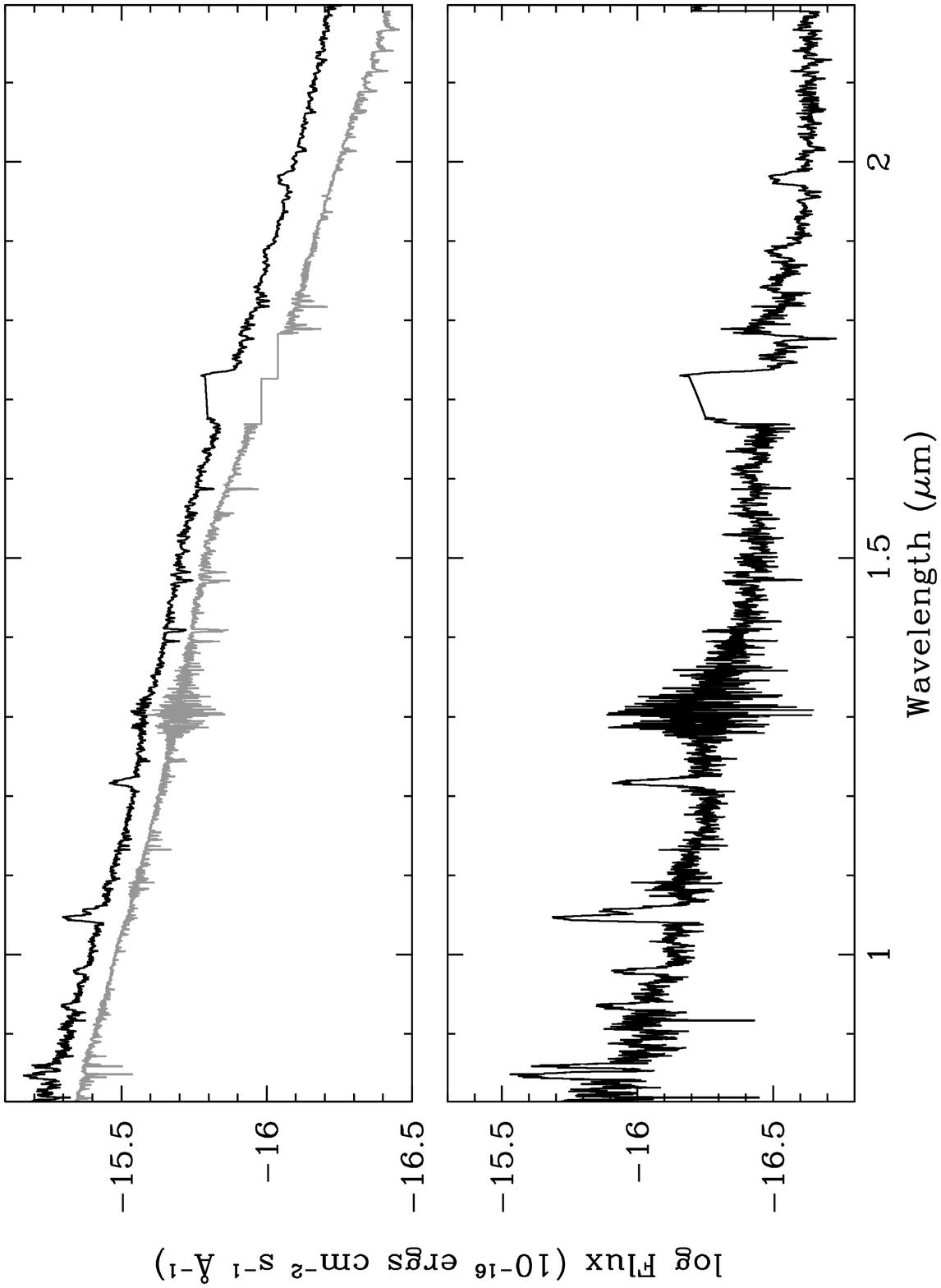}
\centerline{f9.eps}
\end{figure}

\end{document}